# Substrate Sensitive Mid-Infrared Photoresponse in Graphene

Marcus Freitag[1#*], Tony Low[1#], Luis Martin-Moreno[2#], Wenjuan Zhu[1], Francisco Guinea[3], and Phaedon Avouris[1]

[1] IBM T.J. Watson Research Center, Yorktown Heights, NY 10598
[2] Instituto de Ciencia de Materiales de Aragon and Departamento de Fisica de la Materia Condensada, CSIC-Universidad de Zaragoza, E-50009, Zaragoza, Spain
[3] Instituto de Ciencia de Materiales de Madrid. CSIC. Sor Juana Ines de la Cruz 3. 28049 Madrid, Spain

**ABSTRACT:** We report mid-infrared photocurrent spectra of graphene nanoribbon arrays on $SiO_2$ dielectrics showing dual signatures of the substrate interaction. First, hybrid polaritonic modes of graphene plasmons and dielectric surface polar phonons produce a thermal photocurrent in graphene with spectral features that are tunable by gate voltage, nanoribbon width, and light polarization. Secondly, phonon-polaritons associated with the substrate are excited, which indirectly heat up the graphene leading to a graphene photocurrent with fixed spectral features. Models for other commonly used substrates show that the responsivity of graphene infrared photodetectors can be tailored to specific mid-IR frequency bands by the choice of the substrate.

[#] These Authors contributed equally
[*] Corresponding Author email: mfreitag@us.ibm.com

Monolayer 2D systems can interact with the substrate supporting them, and as a result, their properties become substrate-dependent. An interesting case involves the remote interaction between the plasmon modes of graphene[1-6] or graphene nanoribbons[7-10] with the surface polar phonons of a dielectric substrate.[3,9] In particular, photocurrent generation[11] can be strongly affected because the absorption cross-section and the subsequent decay of the excitations, are substrate dependent. Previous measurements of plasmonic photocurrents in graphene nanoribbon arrays[11] involved the excitation of a specific mixed plasmon-phonon mode with a $CO_2$ laser at $943\,\text{cm}^{-1}$ and the photocurrent was shown to modulate as a function of electrostatic doping. In this work, photocurrent spectroscopy of graphene nanoribbon arrays is performed using a quantum cascade laser, allowing the access of a wide range of mid-infrared frequencies from $850\,\text{cm}^{-1}$ to $1600\,\text{cm}^{-1}$. Our study reveals a rich interplay between various polaritonic modes, due to strong coupling between light and various dipole carrying excitations such as plasmons, phonons, and their hybrids. Clear signatures of enhanced light-matter interaction are revealed through the measured spectrally-resolved photocurrent.

Plasmons are collective electronic oscillations and in graphene they follow a square-root dispersion relation $E \sim \sqrt{q}$, where $q$ is the plasmon wave-vector. Direct optical excitation of plasmons in graphene is precluded due to the large momentum mismatch with photons. Near-field excitation is one way to circumvent this.[3-5] In graphene metamaterials, such as arrays of graphene nanoribbons (GNRs), momentum conservation is also relaxed, and standing plasmon modes with momentum that goes with $q \approx \pi/w$, where $w$ is the nanoribbon width, can be excited.[7-9] Not only does the energy of the plasmon depend on the GNR width, it also depends on doping of the graphene. The former makes GNR arrays tunable by design, while the latter makes them *in situ* tunable (over a limited range) by simply applying a back-gate voltage.

For graphene in contact with a polar substrate, interaction of the plasmon with a substrate surface polar phonon (SPP) leads to their hybridization and splitting into two hybrid plasmon-phonon modes. In the case of an $SiO_2$ substrate, the surface polar phonon modes at $460\,\text{cm}^{-1}$, $800\,\text{cm}^{-1}$, and $1170\,\text{cm}^{-1}$ hybridize with the graphene plasmon to form the plasmon-phonon resonances (Fig. 1a). [3, 9] The characteristics of the hybrid modes are inherited partly from plasmons and partly from phonons, and as such, they show dispersion and lifetimes somewhere between the highly dispersive plasmon and the non-dispersive phonon.[9]

In order to detect the optical energy deposited in the GNR array *via* the excitation of a hybrid plasmon-phonon mode, the excitation has to be converted into an electrical signal. A photocurrent in graphene can arise by a number of different mechanisms.[12] Photovoltaic effects,[13-17] thermoelectric effects,[18-21] bolometric effects,[12, 22, 23] and phototransistor action[24, 25] all have been reported. In the photovoltaic effect, built-in electric fields separate photo-generated electrons and holes, which leads to a photocurrent for example upon selective illumination at a contact or illumination at a p-n junction. The thermoelectric effect is caused by the selective light-induced heating of part of the device in combination with spatial variations in the Fermi-level dependent Seebeck coefficient. Phototransistor action requires another material in close proximity where either holes or electrons can get trapped while capacitively coupled to the channel and affecting the transport current. [24, 25] The photocurrent spectrum then reflects the absorption spectrum of the added particles. Finally, in the bolometric effect, biased but otherwise homogeneous graphene is illuminated, which heats it up and changes the transport current.

To explore the plasmonic photocurrents in graphene nanoribbon arrays, we work with homogeneous nanoribbons, since any varying doping profiles such as in p-n junctions would lead to inhomogeneous broadening of the plasmon modes. The photocurrent mechanism under these circumstances is of the bolometric nature, and the sign of the photocurrent depends on the electrostatic doping.[12] Close to the Dirac point, photo-generated carriers dominate and increase the transport current. Away from the Dirac point, the temperature dependence of the carrier mobility dominates, which leads to an overall current decrease. For our p-doped samples the transport current is reduced upon photoexcitation unless we apply a gate voltage on the order of 40V or more. Unless otherwise noted, we focus on the gate-voltage range of -40V to 20V, where the reduction in current due to the heating strongly dominates. Please see Figure 1 and the Methods / Experimental section for the details of the sample and experiment.

**Results / Discussion**

Our tunable Quantum Cascade Laser (QCL) covers mid-infrared frequencies ranging from $850\,\text{cm}^{-1}$ to $1600\,\text{cm}^{-1}$ and therefore allows the excitation of two of the plasmon-phonon modes illustrated in Fig. 1a. Photocurrent spectroscopy (Fig. 2) shows several peaks in this mid-IR window. We identify the most prominent resonances, labeled A and B as two hybrid plasmon-phonon modes[9, 11] due to their clear polarization dependence as shown in Fig. 2a. Electromagnetic simulations of the absorption spectra of arrays of 130nm GNRs, in keeping with the same plasmon oscillator strength $\sqrt{E_F/\varepsilon_{\text{eff}}}$ as in the experiment, are displayed in the top panel of Fig. 2b. The red curve gives the absorption under perpendicular polarization (E-vector perpendicular to the GNR axis) while the blue curve describes the parallel polarization case. The model takes into account the polar optical surface phonons of the $SiO_2$ substrate, which hybridize with graphene plasmons.[3, 8, 9] For the details of the dielectric parameters of $SiO_2$, please see the Supporting Information. The energy of the hybrid plasmon-phonon modes, which are only observed under perpendicular polarization, matches the energy of the experimental photocurrent peaks A and B (Fig. 2a) quite well. Furthermore, the computed absorption in mode A is 3 times larger than the absorption in the mode B, and this is also reflected in the photocurrent spectrum.

Notably absent from the calculated absorption of the GNR array is any mode in the vicinity of $1070\,\text{cm}^{-1}$. This resonance appears in all experimental spectra independently of infrared polarization (Fig. 2a), graphene nanoribbon width, or electrostatic doping (Fig. 3a). In Figure 4 we show an analogous measurement using simple graphene photodetectors (not cut into GNRs), which lacks the plasmon-phonon resonances, but also exhibits this mode at $1070\,\text{cm}^{-1}$. The mode has not been reported before in graphene photocurrent measurements. It is also missing in the absorption spectrum of graphene[26] or graphene nanoribbons[9]. The feature at 1070 cm$^{-1}$ therefore has to have an origin beyond the graphene plasmons or their hybridization with surface polar phonons.

The phonon spectrum of SiO$_2$ includes an infrared active polar phonon near 1100 cm$^{-1}$. When resonantly excited with light, the resultant substrate phonons produce enhanced optical fields at the substrate surface i.e. phonon-polaritons. For a simple semi-infinite SiO$_2$ substrate, normal incidence optical fields decay exponentially according to $\exp(-\mathrm{Im}(k_z)z)$, where $k_z = \sqrt{\varepsilon_{SiO2}}k_0$ and $k_0$ is the free space wave-vector. The light absorption coefficient therefore is proportional to $\exp(-z/l_z)$, where $l_z \equiv 1/2\,\mathrm{Im}(k_z)$ is the absorption depth. The bottom panel of Fig. 2b plots the cumulative absorption from the surface to the depth of 90nm (the SiO$_2$ thickness), showing enhanced surface absorption around 1100 cm$^{-1}$. Fig. 2c plots the intensity of the transverse magnetic field component $|H_y|$ in the device cross section at frequency coinciding with the hybrid plasmon phonon resonance B (1256 cm$^{-1}$) and the infrared-active surface phonon (1112 cm$^{-1}$). It reveals the enhanced field intensity at the interface. The former has surface light confinement that goes with $w/\pi$, while the latter decays with $l_z$, which are both ~100 nm at their respective resonance conditions in our experiment.

The measured bolometric photocurrent is directly proportional to the increase in graphene lattice temperature upon photo-excitation. A simple linear heat-flow model can be applied to estimate the steady state temperature in graphene. Heat flow into the air is several orders smaller than heat flow into the gate stack and we can therefore assume all heat flow into the gate stack. Lateral heat flow along the graphene to the contacts can be neglected since the devices are very long (30μm) compared to the dielectric thickness (90nm). The silicon is assumed to be the heat sink at room temperature, and the temperature drops across the SiO$_2$ with thermal conductivity $\kappa_{SiO2} = 1.5\,\mathrm{Wm^{-1}K^{-1}}$. The thermal contact resistance between graphene and SiO$_2$ is $\kappa_c = 10\,\mathrm{MW/Km^2}$.[27] Laser power is $P=1\mathrm{MW/m^2}$. For absorption in the graphene plasmon-phonon mode, heat generation is a delta function centered at the graphene position, while in the case of the SiO$_2$ phonon, heat is generated continuously along the 90nm dielectric. The former (latter) leads to direct (indirect) heating of graphene. The temperature increase of the graphene given by:

$$\Delta T_{ph} = a_g P \left[ \frac{L}{\kappa_{sio2}} + \frac{1}{\kappa_c} \right]$$

and

$$\Delta T_{ph} = \frac{a_\infty \ell_z P}{\kappa_{sio2}} \left[ 1 - \left(1 + \frac{L}{\ell_z}\right) \exp\left(-\frac{L}{\ell_z}\right) \right]$$

for the direct and indirect heating of the graphene respectively. Here $a_g$ and $a_\infty$ are the simulated light absorption in graphene and semi-infinite dielectric respectively, while in the experiment, a finite dielectric thickness $L$=90nm is used. The calculated temperatures at the position of the graphene as a function of excitation energy are shown in Fig. 2d. The relative intensity of the peaks A, SiO$_2$, and B in Fig. 2a are captured well.

Since the photocurrent measurements with tunable QCL allow us to acquire entire photocurrent spectra, we are now ready to tune the photocurrent maxima by varying GNR widths and electrostatic doping. Figure 3a shows normalized photoconductance spectra for a 90nm GNR array under two different gate voltages. In the zero gate-voltage case (red squares), which corresponds to a Fermi level of $E_F = -0.33\text{eV}$, peaks A and B are slightly up-shifted compared to their counterparts in Figure 2a, which was taken on a 130nm GNR array. Furthermore, at a reduced gate voltage of -40V, which corresponds to $E_F = -0.44\text{eV}$, a strong blue-shift in peaks A and B is observed (blue circles). In addition to the blue-shift, plasmon-phonon mode B broadens substantially with increased electrostatic doping. This broadening is associated with the opening of additional decay-channels for the hybrid plasmon-phonon mode due to optical phonon scattering, which dampens the plasmon.[9] On the other hand, the electrostatic doping does not alter the SiO$_2$ phonon peak, which stays fixed. Finally, the color plots in Figs. 3b,d show the normalized photoconductance in the vicinity of the hybrid plasmon-phonon mode B color-coded as a function of gate voltage and laser energy for both 90nm GNRs and 130nm GNRs. The broadening of peak B, which is very strong for 90nm GNRs is not observed in the 130nm GNR case. In 130nm GNRs, the energy of the hybrid plasmon-phonon mode even with doping at $E_F = -0.44\text{eV}$ is not high enough to reach the energies of the optical phonons that are responsible for the decay channels.

In previous work[9] we have studied the dispersion and damping of these hybrid plasmon-phonon modes within the standard theory based on random phase approximation (RPA), see Supporting Information. Both the substrate phonons and graphene's intrinsic optical phonon are included in this theory. Fig. 3c and 3e plots the RPA plasmon loss function $L$ as a function of frequency and gate voltage for $q$ corresponding to 90nm and 130nm ribbons. Qualitative features of the experiment in Figs. 3b and 3d are captured, including the broadening.

Most dielectrics host vibrational mid-infrared active phonon modes which can also interact with light and plasmons in the same fashion as described above. Here, we consider the cases of hexagonal boron nitride (hBN) and silicon carbide (SiC), common substrates for graphene devices. Their bulk optical phonon frequencies and related dielectric parameters, as well as thermal conductivities are summarized in the Supporting Information. Figure 5a shows the hybrid plasmon-phonon modes for an array of 130nm wide graphene nanoribbons on $SiO_2$, SiC, and hBN, assuming doping of 0.5eV. These various polariton modes distribute across the mid-infrared to the far-infrared, clearly demonstrating that graphene photodetectors can be spectrally tailored by hybridization of the plasmons with substrate phonons. In Fig. 5b we plot the light absorption depth for the various dielectrics as function of frequency, which is responsible for the indirect heating and resulting photocurrent in graphene. Here, we see that silicon carbide and boron nitride both accommodate surface phonon-polaritons, which are more strongly localized than the $SiO_2$ counterpart. However, thermal conductivities of SiC and hBN are 360W/mK and 30W/mK (out-of-plane) respectively, higher than the $SiO_2$ thermal conductivity of 1.5W/mK, which reduces peak temperatures achievable in those materials. Ideally, one would engineer the gate stack in a way that a thin layer of strongly absorbing material such as SiC or BN is deposited onto a thicker dielectric like SiO$_2$ with low thermal conductivity.

**Conclusions**

We presented a spectroscopic study of the photocurrent in graphene nanoribbon arrays over a wide range of mid-IR wavelengths from about 6 μm to 12 μm, (850 cm$^{-1}$ to 1600 cm$^{-1}$). Our experimental observation provides direct proof of the importance of the substrate's phonons in the photocurrent generation process in graphene. These measurements and associated modeling show that graphene photodetection in the mid-infrared can be spectrally tailored in many ways, such as by substrate engineering, designing of the nanoribbon width, and electrostatic doping. Vice versa, graphene photocurrent spectroscopy can reveal signatures of the phononic modes, allowing the vibrational characterization of thin dielectric films or even molecular layers.

**Methods / Experimental**

Our photosensitive graphene structures consist of arrays of graphene nanoribbons 90nm or 130nm in width on an Si/SiO$_2$ substrate (Fig. 1b). The graphene is grown by chemical vapor deposition on copper foil using methane.[28] This process is self-limiting due to the low solubility of carbon in copper, and yields in excess of 95% single-layer graphene with only small patches of few-layer graphene. After depositing PMMA, the copper is dissolved by wet-etching with etchant CE200, and the graphene, now attached to the PMMA, is transferred onto silicon/SiO$_2$ chips with 90nm oxide thickness. With the gate stack in place, source and drain electrodes consisting of 1/20/40 nm Ti/Pd/Au are fabricated by e-beam lithography on top of the graphene. Finally, the graphene is etched into nanoribbons using e-beam lithography, lift-off, and oxygen plasma. In this step, the array dimensions are also established (30 μm long and 10 μm wide).

The as-produced graphene nanoribbon devices are chemically p-doped to a level of 0.33eV as determined from transfer characteristics. Electrostatic doping through the global backgate lets us vary the Fermi level from $E_F = -0.44\text{eV}$ for $V_G = -40\text{V}$ to $E_F = -0.16\text{eV}$ for $V_G = 20\text{V}$. The nanoribbon arrays are designed with GNR width equal to the spacing between them. However by AFM, we measure a GNR width 30nm smaller than the design width, and it is this AFM width that we cite thought this paper. The width as measured by AFM closely matches the electronic width we used to model the plasmons in a previous paper[9], and therefore there is no need for a distinction between electronic and geometric width anymore. The edge roughness of the ribbons is on the order of 10nm, again as measured by AFM.

Transport- and photocurrents are measured in an AC photocurrent setup described in Fig. 1c. Mid-IR radiation from a tunable quantum cascade laser (QCL) is focused into a spot about $20\,\mu\text{m}$ in diameter and centered in the middle of the GNR array. The tuning range of the QCL (Block LaserScope) covers the mid-IR region between $850\,\text{cm}^{-1}$ and $1600\,\text{cm}^{-1}$. The peak laser power varies with wavelength between $5\,\text{mW}$ and $50\,\text{mW}$. Pulses from the QCL have a duty factor of 2.5% at $100\,\text{kHz}$ repetition rate, so that effective AC laser power amplitudes are between $250\,\mu\text{W}$ and $2.5\,\text{mW}$. A bias on the order of -8V is applied at the drain contact of the GNR array, and the source contact is connected to the AC+DC port of a bias Tee. The DC port of the bias Tee is grounded through a preamplifier to measure the transport current, and the AC port is connected to a lock-in amplifier, referenced to the laser pulses at $100\,\text{kHz}$. This allows us to utilize a higher sensitivity preamp for the AC photocurrent measurement on top of the larger DC transport current.

*Conflict of Interest:* The authors declare no competing financial interest.

*Supporting Information Available:* (1) Note concerning the calculation of the RPA loss function and (2) Table specifying the substrate parameters used for the calculations. This material is available free of charge *via* the Internet at http://pubs.acs.org.

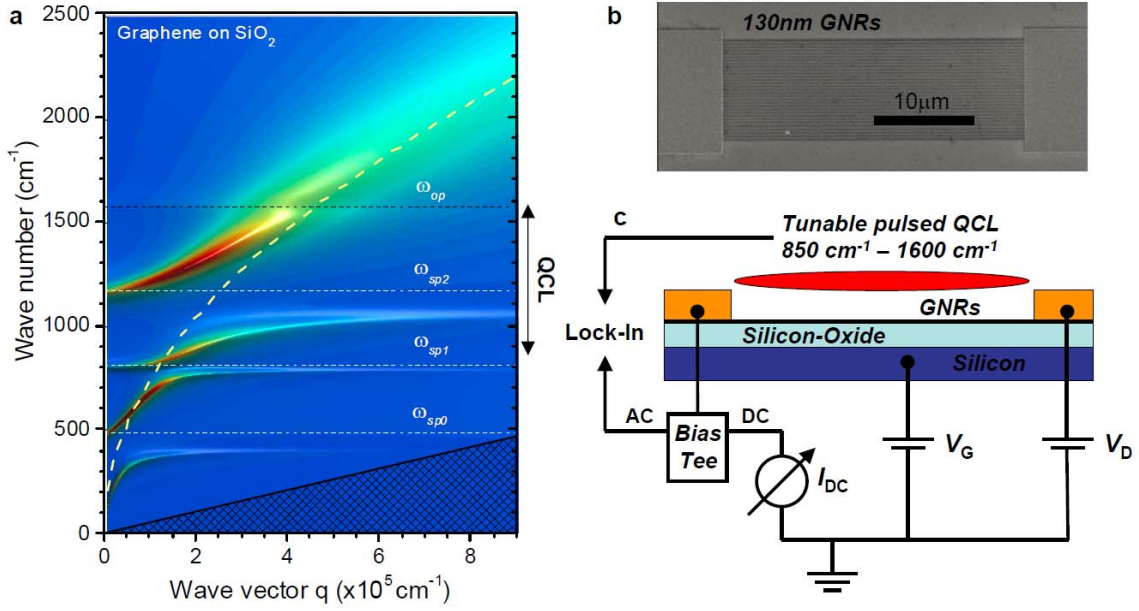

**Figure 1. Mid-IR excitation of GNR array photodetector on Si/SiO$_2$.** (a) Dispersion relation of hybrid plasmon-phonon modes in graphene on SiO$_2$ with chemical potential $\mu = 0.43\,eV$. sp0 to sp2 are polar SiO$_2$ phonons that interact with the graphene plasmon. $\omega_{op}$ is the optical phonon energy in graphene. The dashed curve shows the graphene plasmon dispersion before hybridization with the SiO$_2$ phonon. The shaded triangle indicates the Landau damping region, where plasmons would decay rapidly into e-h pairs. Cutting the graphene into nanoribbons with width $w$ means selecting a wavevector that satisfies the usual half-wavelength condition $q \approx \pi/w$ and therefore choosing specific energies for the resonant plasmon-phonon modes. (b) Contacted array of graphene nanoribbons, 130nm in width and 190nm spacing, fabricated on silicon with 90nm silicon oxide. The entire array is 30μm long and 10μm wide. (c) Mid-IR radiation from a tunable quantum cascade laser, pulsed at 100KHz, is focused by a ZnSe objective into a 20μm spot centered on the GNR array detector. A DC bias on the order of $V_D = -8V$ is applied at the Drain contact. DC and AC electrical signals are separated on the source side by a Bias Tee and sent to either a preamplifier (DC) or a lock-in amplifier (AC) to measure the DC transport current or AC photocurrent respectively.

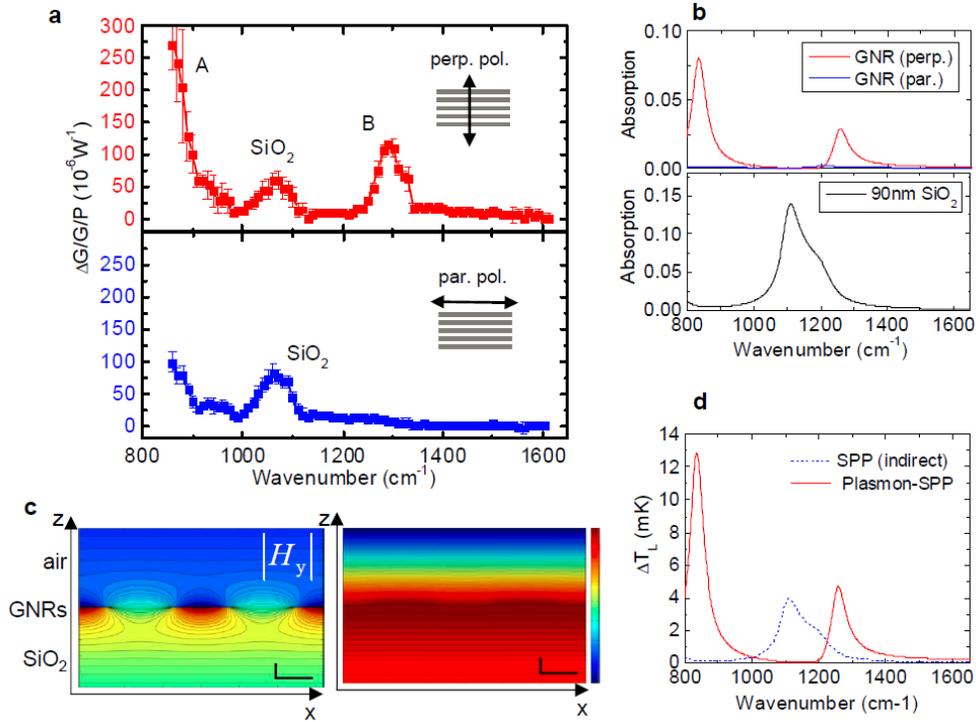

**Figure 2. Photocurrent spectroscopy in the mid-IR.** (**a**) Normalized photoconductance *vs.* excitation energy for an array of 130nm GNRs for two different light polarizations. The photoconductance $\Delta G$ is normalized by the incident laser power $P$ and the dark conductance $G$. Hybrid plasmon-phonon modes A and B are excited and produce a photocurrent for perpendicularly polarized light. The substrate phonon labeled $SiO_2$ on the other hand is not polarization dependent. A residual photocurrent at parallel polarization below peak A is likely due to another infrared-active $SiO_2$ phonon near 800cm$^{-1}$. (**b**) Top panel: Absorption of an array of 130nm GNRs on $SiO_2$ calculated for perpendicular (red) and parallel (blue) polarization. Bottom panel: Calculated absorption due to the infrared-active polar phonon of 90nm $SiO_2$. (**c**) Transverse magnetic field $|H_y|$ contour plots in a plane perpendicular to the GNR array for excitation with energy 1256cm$^{-1}$ in the hybrid plasmon-phonon mode B (left) and with energy 1112cm$^{-1}$ at the $SiO_2$ phonon (right). Scale bars are 100nm. (**d**) Calculated temperature increase at the graphene position upon photoexcitation of the graphene hybrid plasmon-phonon modes (red) and the $SiO_2$ phonon (blue). A $SiO_2$ thermal conductivity of $\kappa_{SiO2} = 1.5$W/mK and interface thermal conductivity between graphene and SiO$_2$ of $\kappa_C = 10\text{MW/m}^2\text{K}$ are used for the calculations.

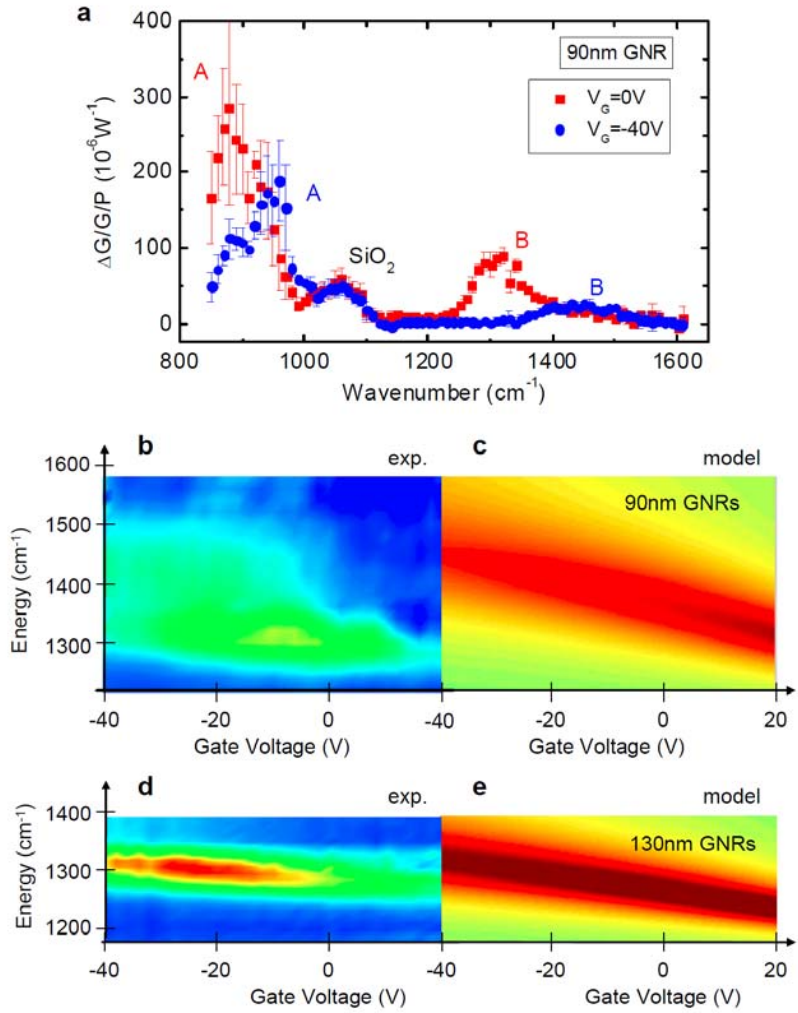

**Figure 3. Doping dependence of the photocurrent spectra.** (a) Normalized photoconductance *vs.* excitation energy for an array of 90nm GNRs and two different backgate voltages. The hybrid plasmon-phonon modes A and B are highly tunable, while the SiO$_2$ phonon is fixed. (b) Plasmon-phonon mode B in 90nm GNRs: The 3D false-color plot shows the experimental photocurrent *vs.* gate voltage and excitation energy. (c) Calculated electron loss function,[11] defined as inverse of the imaginary part of the dielectric function for 90nm GNRs. See the Supporting Information for details of modeling. (d-e) Same as (b-c), but for 130nm GNRs.

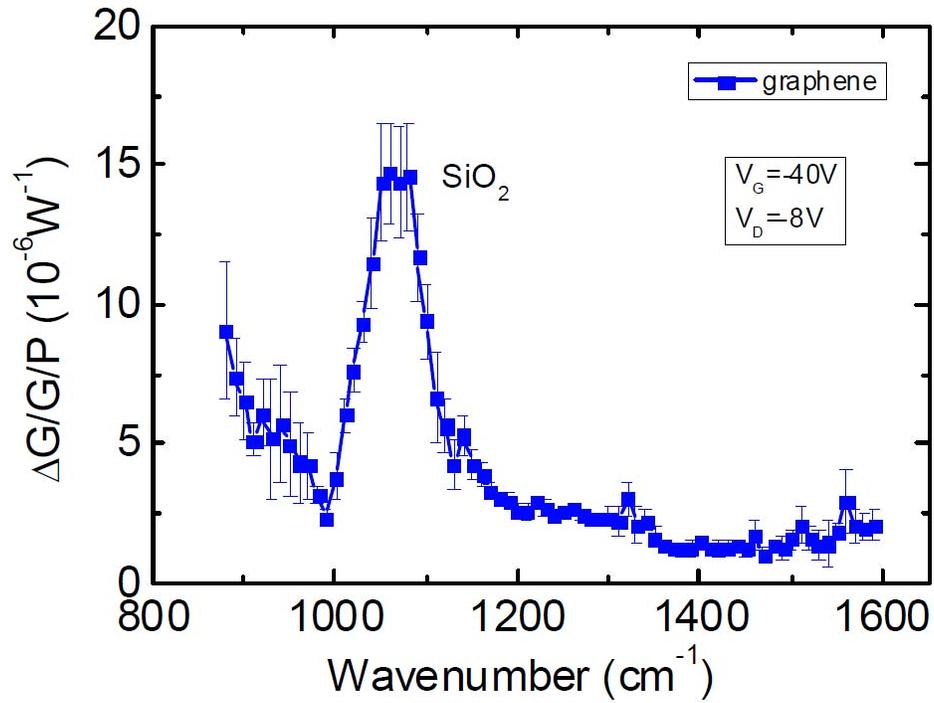

**Figure 4. Photocurrent spectroscopy in the mid-IR for graphene.** The $SiO_2$ related resonance is present even in the absence of plasmonic resonances.

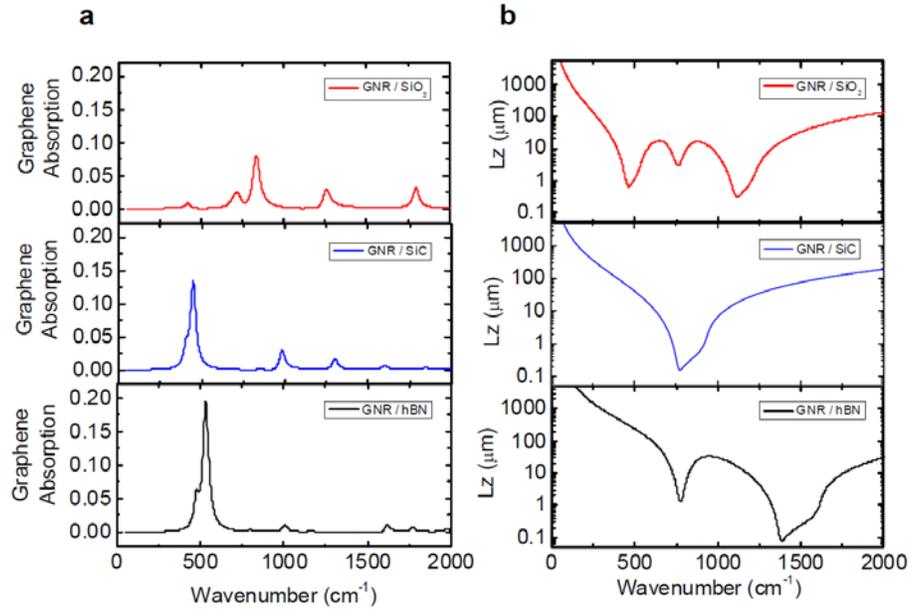

**Figure 5. Effects of various substrate dielectrics. (a)** Absorption in graphene due to plasmon-phonon modes of GNR arrays fabricated on different polar substrates, calculated by solving the Maxwell equation of semi-infinite substrates. A GNR width of 130nm is assumed. **(b)** Light absorption depth for different substrates as indicated. The dielectric parameters for the various substrates are tabulated in the Supporting Information.